\documentclass[showpacs]{revtex4}
\usepackage{amsfonts}
\usepackage{amsmath}
\usepackage{amssymb}

\begin{document}

\title{Thermodynamics on the apparent horizon in generalized gravity theories%
}
\author{Shao-Feng Wu$^{1}$\footnote{%
Email: sfwu@shu.edu.cn}, Bin Wang$^{2}$\footnote{%
Email: wangb@fudan.edu.cn}, and Guo-Hong Yang$^{1}$\footnote{%
Email: ghyang@shu.edu.cn}}
\affiliation{$^{1}$ Department of Physics, Shanghai University, Shanghai 200436, China\\
$^{2}$ Department of Physics, Fudan University, Shanghai 200433, China}

\begin{abstract}
We present a general procedure to construct the first law of thermodynamics
on the apparent horizon and illustrate its validity by examining it in some
extended gravity theories. Applying this procedure, we can describe the
thermodynamics on the apparent horizon in Randall-Sundrum braneworld
imbedded in a nontrivial bulk. We discuss the mass-like function which was
used to link Friedmann equation to the first law of thermodynamics and
obtain its special case which gives the generalized Misner-Sharp mass in
Lovelock gravity.
\end{abstract}

\maketitle

\section{Introduction}

Inspired by black hole thermodynamics, it was realized that there is a
profound connection between gravity and thermodynamics. Jacobson first
showed that the Einstein gravity can be derived from the first law of
thermodynamics in the Rindler spacetime. For a general static spherically
symmetric spacetime, Padmanabhan pointed out that Einstein equations at the
horizon give rise to the first law of thermodynamics \cite{Padmanabhan}.
Recently the study on the connection between gravity and thermodynamics has
been extended to cosmological context. Frolov and Kofman \cite{Frolov}
employed the approach proposed by Jacobson \cite{Jacobson} to a quasi-de
Sitter geometry of inflationary universe, and calculated the energy flux of
a background slow-roll scalar through the quasi-de Sitter apparent horizon.
By applying the first law of thermodynamics to a cosmological horizon,
Danielsson obtained Friedmann equation in the expanding universe \cite%
{Danielsson}. In the quintessence dominated accelerating universe, Bousso
\cite{Bousso} showed that the first law of thermodynamics holds at the
apparent horizon. Cai and Kim \cite{Cai} generalized the derivation of the
Friedmann equations from the first law of thermodynamics to the spacetime
with any spatial curvature. This study has also been generalized to the $%
f(R) $ gravity \cite{Eling,Akbar} and scalar-tensor gravity theory \cite{Cao}%
..

Besides gravity theories in four dimensions, the study on the connection
between gravity and thermodynamics has also been extended to the braneworld
cosmology \cite{Cao1,Sheykhi,Sheykhi1,Wu1}. It has been suggested that the
first law of thermodynamics on the apparent horizon can be derived from the
Friedmann equations in the Randall-Sundrum braneworld \cite{Cao1,Sheykhi},
also in braneworld with curvature corrections including 4D scalar curvature
from induced gravity on the brane and the 5D Gauss-Bonnet curvature
correction \cite{Sheykhi1}, and the braneworld with finite brane thickness
\cite{Wu1}. In the braneworld the exact black hole solution has not been
discovered until now. It was pointed out that the connection between gravity
and thermodynamics can shed lights on the entropy of the braneworld \cite%
{Sheykhi,Sheykhi1}.

It is still unclear whether the connection between gravity and
thermodynamics also holds in more general gravity theories. It was argued
that if one does not know the bulk geometry, it is not easy to build up
thermodynamics in such spacetime \cite{Cao1}. However, if one looks at $f(R)$
gravity, although the concrete form in this nonlinear gravity theory is not
available, the corresponding thermodynamics can still be obtained, with the
price of introducing the entropy production. In this work we are going to
investigate this problem. We will present a general procedure to construct
thermodynamics on the apparent horizon and show that this procedure
correctly works, examined in some extended gravity theories. We will try to
explain the reason that the additional entropy production terms appear in
the $f(R)$ gravity and scalar-tensor gravity theories, while they do not
appear in Einstein and Lovelock gravity theories etc. With the general
procedure, we will study the thermodynamics in the Randall-Sundrum
braneworld imbedded in the nontrivial bulk. Recently, the model has been
used to mimic the phantom-like dark energy without violating weak energy
condition \cite{Cai3,Apostolopoulos,wang1,wang2}.

In \cite{Gong}, it was argued that there is a mass-like function connecting
the first law of thermodynamics and the Friedmann equations in some gravity
theories. It is of interest to investigate whether this mass-like function
has the connection with the Misner-Sharp mass. The Misner-Sharp mass is
widely accepted as a quasi-local mass in Einstein gravity and its
generalized forms have been obtained in Gauss-Bonnet gravity, Lovelock
gravity, etc \cite{Maeda,Maeda1}. We will show that in Lovelock gravity, the
generalized Misner-Sharp mass is a special form of the mass-like function.

The organization of the paper is the following: In section 2, we will
present the general procedure to construct the first law of thermodynamics
on the apparent horizon. In the following section we will examine this
procedure and show that it is valid by illustrating it to some extended
gravity theories including Lovelock, $f(R)$ and scalar-tensor gravity
theories. In section 4, we will apply the general procedure to study the
thermodynamics on the apparent horizon in Randall-Sundrum braneworld
imbedded in the nontrivial bulk. Our conclusion and discussion will be
present in the last section.

\section{The first law of thermodynamics on apparent horizon of FRW cosmology%
}

In this section, we are going to present a general procedure to construct
the first law of thermodynamics on the apparent horizon. We will consider
extended gravity theories based on Einstein gravity.

The homogenous and isotropic ($n+1$)-dimensional FRW universe is described by%
\begin{equation}
ds^{2}=h_{ab}dx^{a}dx^{b}+\tilde{r}^{2}d\Omega_{n-1}^{2},  \label{FRW}
\end{equation}
where $h_{ab}=$diag$(-1,\frac{a^{2}}{1-ka^{2}})$, $d\Omega_{n-1}^{2}$ is the
$\left( n-1\right) $-dimensional sphere element, and $x^{0}=t,\;x^{1}=r,\;%
\tilde{r}=ar$ is the radius of the sphere and $a$ is the scale factor. For
simplicity, we consider the flat space $k=0$ in this paper, however, our
discussion can also be generalized to the non-flat cases. It is known that
the dynamical apparent horizon, the marginally trapped surface with
vanishing expansion, is defined as a sphere situated at $r=r_{A}$ satisfying%
\begin{equation*}
h^{ab}\partial_{a}\tilde{r}\partial_{b}\tilde{r}=0.
\end{equation*}
The sphere radius is%
\begin{equation}
\tilde{r}_{A}\equiv r_{A}a=\frac{1}{H}.  \label{Horizon}
\end{equation}
The associated temperature on the apparent horizon can be defined as%
\begin{equation}
T=\frac{1}{2\pi\tilde{r}_{A}}.  \label{T}
\end{equation}
In Einstein gravity, the entropy is proportional to the horizon area%
\begin{equation*}
S_{E}=\frac{A}{4G},
\end{equation*}
where the horizon area $A=n\Omega_{n}\tilde{r}_{A}^{n-1}$, thus we have the
fundamental relation%
\begin{equation*}
\delta Q\equiv TdS_{E}=\frac{n(n-1)V\tilde{r}_{A}^{-3}d\tilde{r}_{A}}{8\pi G}%
,
\end{equation*}
where $V=\Omega_{n}\tilde{r}_{A}^{n}$ is the volume in the horizon. Using
the definition of the horizon (\ref{Horizon}), we can obtain%
\begin{equation}
TdS_{E}=\frac{-n(n-1)V}{16\pi G}\frac{dH^{2}}{dt}dt,  \label{dr1}
\end{equation}
which is purely a geometric relation.

For all gravity theories, one can write the Friedmann equations as the form
of Einstein gravity%
\begin{equation}
H^{2}=\frac{16\pi G}{n(n-1)}\rho _{eff}  \label{H21}
\end{equation}%
\begin{equation}
\dot{H}=-\frac{8\pi G}{(n-1)}(\rho _{eff}+p_{eff}).  \label{H22}
\end{equation}%
Though we do not know the exact form of $\rho _{eff}$ (and $p_{eff}$), we
know that there must be ordinary matter density $\rho $ in $\rho _{eff}$ and
also other quantities $\rho _{i}$, such as matter or energy components
besides the ordinary matter. The first Friedmann equation can be expressed
in the form%
\begin{equation*}
H^{2}=H^{2}(\rho ,\;\rho _{1},\cdots \rho _{i},\cdots ).
\end{equation*}%
Then the relation (\ref{dr1}) can be changed as%
\begin{equation}
TdS_{E}=\frac{-n(n-1)V}{16\pi G}dt(\frac{\partial H^{2}}{\partial \rho }\dot{%
\rho}+\frac{\partial H^{2}}{\partial \rho _{i}}\dot{\rho}_{i}).  \label{dr2}
\end{equation}%
To construct the first law of thermodynamics $dE=TdS$, we need to know the
energy flux $dE$ and entropy $S$. In the general gravity theory, they are
not specified. The energy flux of ordinary matter can be expressed as $dE=V%
\dot{\rho}dt$. Multiplying $\frac{16\pi G}{n(n-1)}\frac{1}{\frac{\partial
H^{2}}{\partial \rho }}$ on both sides of (\ref{dr2}), we have%
\begin{equation}
\frac{16\pi G}{n(n-1)}\frac{1}{\frac{\partial H^{2}}{\partial \rho }}%
TdS_{E}=-V\dot{\rho}dt-Vdt\frac{1}{\frac{\partial H^{2}}{\partial \rho }}%
\frac{\partial H^{2}}{\partial \rho _{i}}\dot{\rho}_{i}.  \label{dr3}
\end{equation}%
In the general case, the conservation of the total matter density can be
written as%
\begin{equation}
\dot{\rho}_{eff}+nH(\rho _{eff}+p_{eff})=0,  \label{Con1}
\end{equation}%
while we assume that ordinary matter has energy exchange $q$ with other
matter or energy contents,%
\begin{equation}
\dot{\rho}+nH(\rho +p)=q.  \label{Con2}
\end{equation}%
Equations (\ref{Con1}) and (\ref{Con2}) will be used to express the first
law explicitly. Substituting (\ref{Con2}), (\ref{dr3}) can be changed to the
form
\begin{equation}
T\frac{16\pi G}{n(n-1)}\frac{1}{\frac{\partial H^{2}}{\partial \rho }}%
dS_{E}=nVH(\rho +p)dt-Vqdt-Vdt\frac{1}{\frac{\partial H^{2}}{\partial \rho }}%
\frac{\partial H^{2}}{\partial \rho _{i}}\dot{\rho}_{i}.  \label{dr4}
\end{equation}%
The entropy form can be got by integrating (\ref{dr4}). If there is just
ordinary matter $\rho $ in the space, $\frac{\partial H^{2}}{\partial \rho }$
can be rewritten as a function of $\tilde{r}_{A}$. Then the entropy can be
obtained by the integration%
\begin{align}
S& =\int \frac{16\pi G}{n(n-1)}\frac{1}{\frac{\partial H^{2}}{\partial \rho }%
(\tilde{r}_{A})}d\left( S_{E}\right)  \notag \\
& =\int \frac{4\pi \tilde{r}_{A}^{n-2}\Omega _{n}}{\frac{\partial H^{2}}{%
\partial \rho }(\tilde{r}_{A})}d\tilde{r}_{A}  \label{S0}
\end{align}%
and the relation (\ref{dr4}) can be written as%
\begin{equation}
TdS=dE,  \label{first law1}
\end{equation}%
where%
\begin{equation*}
dE=V\dot{\rho}dt=nVH(\rho +p)dt.
\end{equation*}%
It is the first law of thermodynamics for the gravity theories with only
freedom $\rho $ in the first Friedmann equation.

If the extended gravity theory has other dynamic fields resulting that $%
\frac{\partial H^{2}}{\partial \rho }$ is a function of $\tilde{r}_{A}$ and $%
\rho _{i}$%
\begin{equation*}
\frac{\partial H^{2}}{\partial \rho }=\frac{\partial H^{2}}{\partial \rho }(%
\tilde{r}_{A},\rho _{i}).
\end{equation*}%
The last term on the r.h.s in Eq. (\ref{dr4}) can not be included in the
total differential in general. However we can express the l.h.s in (\ref{dr4}%
) in the form
\begin{equation}
T\frac{16\pi G}{n(n-1)}\frac{1}{\frac{\partial H^{2}}{\partial \rho }}%
dS_{E}=Td\left( \frac{16\pi G}{n(n-1)}\frac{1}{\frac{\partial H^{2}}{%
\partial \rho }}S_{E}\right) -T\frac{16\pi G}{n(n-1)}S_{E}d\frac{1}{\frac{%
\partial H^{2}}{\partial \rho }}.  \label{first law20}
\end{equation}%
Then the general expression of the first law of thermodynamics for
generalized gravity theories with more freedoms in the first Friedmann
equation reads%
\begin{equation}
TdS+Td_{i}S=dE.  \label{first law2}
\end{equation}%
$d_{i}S$ is defined as%
\begin{equation}
d_{i}S\equiv -\frac{16\pi G}{n(n-1)}S_{E}d\frac{1}{\frac{\partial H^{2}}{%
\partial \rho }}=-\frac{4\pi \tilde{r}_{A}^{n-1}\Omega _{n}}{(n-1)}d\frac{1}{%
\frac{\partial H^{2}}{\partial \rho }},  \label{dis}
\end{equation}%
which is interesting since it relates to the entropy production in the
nonequilibrium thermodynamics. This entropy production term comes because it
cannot be absorbed in the complete derivative as in the usual gravity
theory. The energy flux is defined as
\begin{equation}
dE\equiv -V\dot{\rho}dt-Vdt\frac{1}{\frac{\partial H^{2}}{\partial \rho }}%
\frac{\partial H^{2}}{\partial \rho _{i}}\dot{\rho}_{i}=nVH(\rho
+p)dt-Vqdt-Vdt\frac{1}{\frac{\partial H^{2}}{\partial \rho }}\frac{\partial
H^{2}}{\partial \rho _{i}}\dot{\rho}_{i}.  \label{dE}
\end{equation}%
The entropy can be obtained as
\begin{equation}
S=\frac{16\pi G}{n(n-1)}\frac{1}{\frac{\partial H^{2}}{\partial \rho }}S_{E}=%
\frac{4\pi \tilde{r}_{A}^{n-1}\Omega _{n}}{(n-1)}\frac{1}{\frac{\partial
H^{2}}{\partial \rho }}.  \label{S}
\end{equation}%
The exact form of entropy production $d_{i}S$, energy flux $dE$ and entropy $%
S$ depend on the corresponding gravity theory.

\bigskip Recently, a generalized mass-like function in ($3+1$)-dimensional
Einstein gravity%
\begin{equation}
M\equiv\frac{\tilde{r}}{2G}(1+h^{ab}\tilde{r}_{,a}\tilde{r}_{,b})  \label{M0}
\end{equation}
has been used to connect the Friedmann equations and the first law of
thermodynamics on the apparent horizon \cite{Gong}
\begin{equation}
TdS_{E}=\frac{1}{G}d\tilde{r}_{A}=k^{a}\nabla_{a}Mdt=4\pi\tilde{r}%
_{A}^{3}H(\rho+p)dt=dE.  \label{M}
\end{equation}
The mass formulae%
\begin{equation*}
M_{,a}=-4\pi\tilde{r}^{2}(T_{a}^{b}-\delta_{a}^{b}T)\tilde{r}_{,b}+\tilde {r}%
_{,a}
\end{equation*}
has been used to derive the third equality in (\ref{M}). But whether this
mass-like function is general and valid is not clear. Here we will address
this point.

The essence of relation (\ref{M}) is to introduce the $M$ satisfying the
geometric relation%
\begin{equation*}
TdS_{E}=k^{a}\nabla _{a}Mdt
\end{equation*}%
where $k^{a}$ is the\ (approximate) generator $k^{a}=(1,-rH)$ of the
apparent horizon which is null at the horizon. The third equality of
relation (\ref{M}) can be proved by Friedmann equations (\ref{H21}) and (\ref%
{H22}), even without introducing the mass formulae in (\ref{M}). To derive
the generalized mass-like function, one must know the entropy expression at
first. Substituting the entropy expression (\ref{S0}) we have equalities on
the horizon%
\begin{equation}
TdS=\frac{16\pi G}{n(n-1)}\frac{1}{\frac{\partial H^{2}}{\partial \rho }(%
\tilde{r}_{A})}TdS_{E}=\frac{16\pi G}{n(n-1)}\frac{1}{\frac{\partial H^{2}}{%
\partial \rho }(\tilde{r}_{A})}k^{a}\nabla _{a}\left[ M+f_{1}\right]
dt=k^{a}\nabla _{a}\tilde{M}dt,  \label{me}
\end{equation}%
where $M$ is the ($n+1$)-dimensional generalization of the mass-like
function (\ref{M0}) in ($3+1$) dimensions%
\begin{equation*}
M=\frac{n(n-1)\Omega _{n}\tilde{r}^{n-2}}{16\pi G}(1+h^{ab}\tilde{r}_{,a}%
\tilde{r}_{,b}).
\end{equation*}%
Noting $\frac{\partial H^{2}}{\partial \rho }$ is just a function of $\tilde{%
r}_{A}$, using the constraint%
\begin{equation}
k^{a}\tilde{r}_{,a}=0,  \label{km0}
\end{equation}%
and remembering that $f_{i}$ ($i=1,2$) are arbitrary functions satisfying%
\begin{equation}
k^{a}f_{i,a}=0  \label{fi}
\end{equation}%
on the horizon, we can get the last equality of (\ref{me}) provided that%
\begin{equation}
\tilde{M}=\frac{16\pi G}{n(n-1)}\frac{1}{\frac{\partial H^{2}}{\partial \rho
}}\left[ M+f_{1}\right] +f_{2}.  \label{mbar}
\end{equation}%
The mass-like function $\tilde{M}$ is determined up to functions $f_{i}$,
which means that $\tilde{M}$ can have many desired properties in different
cases by tuning the functions $f_{i}$. For the most simplest case, the
Einstein gravity, if we set $f_{1}=-\frac{2n(n-1)\Omega _{n}\tilde{r}%
_{A}^{n-2}}{16\pi G}$ and $f_{2}=0$, the mass function $\tilde{M}$ is just
the negative Misner-Sharp mass
\begin{equation*}
\tilde{M}=-M_{ms}=\frac{n(n-1)\Omega _{n}\tilde{r}^{n-2}}{16\pi G}(-1+h^{ab}%
\tilde{r}_{,a}\tilde{r}_{,b}).
\end{equation*}%
The situation is different for the gravity theories with more matter or
energy components. Using the entropy expression (\ref{S}), we have the
following equation at the horizon
\begin{align*}
TdS& =\frac{16\pi G}{n(n-1)}\frac{1}{\frac{\partial H^{2}}{\partial \rho }(%
\tilde{r}_{A},\rho _{i}(t))}TdS_{E}+\frac{n\Omega _{n}\tilde{r}_{A}^{n-2}}{%
8\pi G}d(\frac{16\pi G}{n(n-1)}\frac{1}{\frac{\partial H^{2}}{\partial \rho }%
(\tilde{r}_{A},\rho _{i}(t))}) \\
& =\frac{16\pi G}{n(n-1)}\frac{1}{\frac{\partial H^{2}}{\partial \rho }(%
\tilde{r}_{A},\rho _{i}(t))}k^{a}\nabla _{a}\left( M+f_{3}\right) dt+\frac{2%
}{(n-1)}\left( M+f_{3}\right) d(\frac{16\pi G}{n(n-1)}\frac{1}{\frac{%
\partial H^{2}}{\partial \rho }(\tilde{r}_{A},\rho _{i}(t))}) \\
& =k^{a}\nabla _{a}\tilde{M}dt,
\end{align*}%
where $f_{3}$ satisfies%
\begin{equation*}
k^{a}f_{3,a}=0,
\end{equation*}%
and $f_{3}=0$ at the horizon. Noting now $\frac{\partial H^{2}}{\partial
\rho }$ is a function of $\tilde{r}_{A}$ and $\rho _{i}$, and using the
constraint (\ref{km0}), we finds that only when $n=3$, the last equation has
the general solution
\begin{equation}
\tilde{M}=\frac{16\pi G}{n(n-1)}\frac{1}{\frac{\partial H^{2}}{\partial \rho
}}\left( M+f_{3}\right) +f_{2},\;n=3.  \label{mbar1}
\end{equation}%
Setting $f_{1}=f_{3}$, expressions of mass-like function in (\ref{mbar}) and
(\ref{mbar1}) are the same when $n=3$. The mass-like function $\tilde{M}$
embodies the specific property in the $(3+1)$-dimensional gravity.

\section{Thermodynamics of some extended gravity theories}

In this section, we will calculate the entropy expressions (\ref{S0}) (\ref%
{S}) and give explicitly the first law of thermodynamics (\ref{first law1}) (%
\ref{first law2}) in some extended gravity theories, including the Lovelock,
$f(R)$ and scalar-tensor gravity theories. We will give explicitly the
mass-like functions accordingly.

\subsection{Lovelock gravity}

The Lagrangian of the Lovelock gravity \cite{Lovelock} consists of the
dimensionally extended Euler densities%
\begin{equation*}
L=\sum_{i=1}^{[n/2]}c_{i}L_{i},
\end{equation*}
where $c_{i}$ is an arbitrary constant and $L_{i}$ is the Euler density of a
($2i$)-dimensional manifold%
\begin{equation*}
L_{i}=2^{-i}\delta_{\alpha_{1}\beta_{1}\cdots\alpha_{ii}\beta_{i}}^{\mu_{1}%
\nu_{1}\cdots\mu_{i}\nu_{i}}R_{\mu_{1}\nu_{1}\cdots\mu_{i}\nu_{i}}^{\alpha
_{1}\beta_{1}\cdots\alpha_{ii}\beta_{i}}.
\end{equation*}
$L_{1}$ is just the Einstein-Hilbert term, and $L_{2}$ corresponds to the so
called Gauss-Bonnet term. Using the FRW metric, we obtain the Friedmann
equations in ($n+1$)-dimensional space-time%
\begin{equation}
\sum_{i=1}^{[n/2]}\hat{c}_{i}\left( H^{2}\right) ^{i}=\frac{16\pi G}{n(n-1)}%
\rho,  \label{FM1Lovelock}
\end{equation}
and%
\begin{equation*}
\sum_{i=1}^{[n/2]}\hat{c}_{i}i\left( H^{2}\right) ^{i-1}(\dot{H})=-\frac{%
8\pi G}{(n-1)}(\rho+p),
\end{equation*}
where%
\begin{equation*}
\hat{c}_{i}=\frac{(n-2)!}{(n-2i)!}c_{i}
\end{equation*}
Since there is only one freedom $\rho$ in the first Friedmann equation (\ref%
{FM1Lovelock}), we will use the entropy expression (\ref{S0}), and obtain%
\begin{align}
S & =\int\frac{4\pi\tilde{r}_{A}^{n-2}\Omega_{n}}{\frac{\partial H^{2}}{%
\partial\rho}(\tilde{r}_{A})}d\tilde{r}_{A}  \notag \\
& =\frac{n\Omega_{n}}{4G}\sum_{i=1}^{[n/2]}\frac{i(n-1)}{n-2i+1}\hat{c}_{i}%
\tilde{r}_{A}^{n-2i+1}.  \label{S lovelock}
\end{align}
and the corresponding first law (\ref{first law1}) reads%
\begin{equation*}
TdS=dE=n\Omega_{n}\tilde{r}_{A}^{n}H(\rho+p),
\end{equation*}
where the second equality holds since there is not $q$ and $\rho_{i}$.

\bigskip The mass-like function (\ref{mbar}) now reads%
\begin{align}
\tilde{M} & =\frac{16\pi G}{n(n-1)}\frac{1}{\frac{\partial H^{2}}{%
\partial\rho}(\tilde{r}_{A})}\left[ M+f_{1}\right] +f_{2}  \notag \\
& =\frac{n(n-1)\Omega_{n}\tilde{r}_{A}^{n}}{16\pi G}(1+h^{ab}\tilde{r}_{,a}%
\tilde{r}_{,b})\sum_{i=1}^{[n/2]}\hat{c}_{i}i\tilde{r}_{A}^{-2i}+f_{1}%
\sum_{i=1}^{[n/2]}\hat{c}_{i}i\tilde{r}_{A}^{-2i}+f_{2}.  \label{mbar love}
\end{align}
In Ref. \cite{Maeda}, the generalized Misner-Sharp mass for Lovelock gravity
was conjectured
\begin{equation}
M_{ms}=\frac{n(n-1)\Omega_{n}\tilde{r}_{A}^{n}}{16\pi G}\sum_{i=1}^{[n/2]}%
\hat{c}_{i}\tilde{r}_{A}^{-2i}(1-h^{ab}\tilde{r}_{,a}\tilde{r}_{,b})^{i}.
\label{MSLovelock}
\end{equation}
Setting%
\begin{equation*}
f_{1}=-2\frac{n(n-1)\Omega_{n}\tilde{r}_{A}^{n}}{16\pi G}
\end{equation*}%
\begin{equation*}
f_{2}=\frac{n(n-1)\Omega_{n}\tilde{r}_{A}^{n}}{16\pi G}\sum_{i=1}^{[n/2]}%
\hat{c}_{i}\tilde{r}_{A}^{-2i}\left[ i(1-h^{ab}\tilde{r}_{,a}\tilde{r}%
_{,b})-(1-h^{ab}\tilde{r}_{,a}\tilde{r}_{,b})^{i}\right] ,
\end{equation*}
which satisfy (\ref{fi}), we can easily find that $M_{ms}$ in (\ref%
{MSLovelock}) is the same as negative (\ref{mbar love}). Thus, the
generalized Misner-Sharp mass can be considered as a specific case of the
(negative) mass function $\tilde{M}$.

\subsection{ Nonlinear gravity}

\bigskip For the nonlinear gravity $f(R)$, the Lagrangian is%
\begin{equation*}
L=\frac{1}{16\pi G}f(R)
\end{equation*}%
The variational principle gives equations of motion. Using the FRW metric,
one can obtain the Friedmann equations in ($n+1$)-dimensional space-time%
\begin{equation}
H^{2}=\frac{16\pi G}{n(n-1)}\frac{1}{f^{\prime }}\left( \rho +\rho
_{c}f^{\prime }\right)  \label{FMfr1}
\end{equation}%
\begin{equation}
\dot{H}=-\frac{8\pi G}{(n-1)}\frac{1}{f^{\prime }}(\rho +\rho _{c}f^{\prime
}+p+p_{c}f^{\prime }),  \label{FMfr2}
\end{equation}%
where%
\begin{equation*}
\rho _{c}=\frac{1}{8\pi Gf^{\prime }}\left[ -\frac{f-Rf^{\prime }}{2}%
-nHf^{\prime \prime }\dot{R}\right]
\end{equation*}%
\begin{equation*}
p_{c}=\frac{1}{8\pi Gf^{\prime }}\left[ (f-Rf^{\prime })-f^{\prime \prime }%
\ddot{R}+f^{\prime \prime \prime }\dot{R}^{2}+n(n-1)f^{\prime \prime }\dot{R}%
\right] .
\end{equation*}%
There are other freedoms besides $\rho $ in the first Friedmann equation (%
\ref{FMfr1}), so one should consider the non-equilibrium thermodynamics. One
can select $\rho _{i}$ arbitrarily. For example, we select $\rho
_{i}=(f^{\prime },\rho _{c})$. But it should be noticed that there is not
real matter fields besides the ordinary matter $\rho $. From the first
Friedmann equation (\ref{FMfr1}), one can explain $\rho _{e}\equiv \rho
_{c}f^{\prime }$ ($p_{e}\equiv p_{c}f^{\prime }$) as the density (pressure)
of an effective energy component in $f(R)$ gravity. Now recalling the
conserved equation (\ref{Con1})%
\begin{equation}
\dot{\rho}_{eff}=-nH(\rho _{eff}+p_{eff})  \label{con}
\end{equation}%
and using the first Friedmann equation (\ref{FMfr1}), one can find that the
l.h.s in Eq. (\ref{con}) reads
\begin{align}
\dot{\rho}_{eff}& =\frac{n(n-1)}{16\pi G}\left( \frac{\partial H^{2}}{%
\partial \rho }\dot{\rho}+\frac{\partial H^{2}}{\partial \rho _{i}}\dot{\rho}%
_{i}\right)  \notag \\
& =\frac{n(n-1)}{16\pi G}\left( \frac{16\pi G}{n(n-1)}\frac{1}{f^{\prime }}%
\dot{\rho}+\frac{\partial H^{2}}{\partial \rho _{i}}\dot{\rho}_{i}\right)
\notag \\
& =\frac{1}{f^{\prime }}\dot{\rho}+\frac{n(n-1)}{16\pi G}\frac{\partial H^{2}%
}{\partial \rho _{i}}\dot{\rho}_{i}  \label{r1}
\end{align}%
while the r.h.s in Eq. (\ref{con}) reads by the second Friedmann equation (%
\ref{FMfr2})
\begin{equation}
-nH(\rho _{eff}+p_{eff})=-nH\left[ \frac{1}{f^{\prime }}\left( \rho +p+\rho
_{e}+p_{e}\right) \right] .  \label{r2}
\end{equation}%
Employing the continuous equation (\ref{Con2}) to Eqs. (\ref{r1}) and (\ref%
{r2}), one can find%
\begin{equation}
\frac{\partial H^{2}}{\partial \rho _{i}}\dot{\rho}_{i}=-\frac{16\pi G}{(n-1)%
}\frac{1}{f^{\prime }}\left[ H\left( \rho _{e}+p_{e}\right) +q\right] .
\label{ls}
\end{equation}%
By substituting Eq. (\ref{ls}) into Eq. (\ref{dE}), we have%
\begin{align}
dE& =nVH(\rho +p)dt-Vqdt-Vdt\frac{1}{\frac{\partial H^{2}}{\partial \rho }}%
\frac{\partial H^{2}}{\partial \rho _{i}}\dot{\rho}_{i}  \notag \\
& =nVH(\rho +p)dt+nVH(\rho _{e}+p_{e})dt.  \label{frdE}
\end{align}%
Then we have the first law%
\begin{equation*}
TdS+Td_{i}S=dE,
\end{equation*}%
where the entropy production and entropy can be obtained from Eq. (\ref{dis}%
) and Eq. (\ref{S}) respectively
\begin{equation}
d_{i}S=-S_{E}df^{\prime },  \label{dis1}
\end{equation}%
and%
\begin{equation}
S=S_{E}f^{\prime }=\frac{A}{4G}f^{\prime }.  \label{Sfr}
\end{equation}%
It should be noticed that the energy flux in Eq. (\ref{frdE}) is different
with the defination in \cite{Eling,Akbar}. But one can find that it is
indeed originated from the energy flux of matter and effective energy
component in $f(R)$ gravity.

Setting $f_{2}=f_{3}=0$, the mass-like function (\ref{mbar1}) for $n=3$ is%
\begin{align}
\tilde{M} & =\frac{16\pi G}{n(n-1)}\frac{1}{\frac{\partial H^{2}}{%
\partial\rho}}M  \label{mbar fr} \\
& =\frac{n(n-1)\Omega_{n}\tilde{r}_{A}^{n-2}}{16\pi G}(1+h^{ab}\tilde{r}_{,a}%
\tilde{r}_{,b})f^{\prime}.  \notag
\end{align}

\subsection{Scalar-tensor gravity}

The general scalar-tensor theory of gravity is described by the Lagrangian%
\begin{equation*}
L=\frac{1}{16\pi G}F\left( \phi \right) R-\frac{1}{2}g^{\mu \nu }\partial
_{\mu }\phi \partial _{\nu }\phi -V\left( \phi \right) ,
\end{equation*}%
where $F(\phi )$ is a positive continuous function of the scalar field $\phi
$ and $V(\phi )$ is its potential. Using the FRW metric, we obtain the
Friedmann equations in ($n+1$)-dimensional space-time%
\begin{equation}
H^{2}=\frac{16\pi G}{n(n-1)}\frac{1}{F}\left( \rho +\rho _{f}+\rho
_{c}F\right)  \label{FMfai1}
\end{equation}%
\begin{equation}
\dot{H}=\frac{8\pi G}{(n-1)}\frac{1}{F}\left( \rho +p+\rho _{f}+p_{f}+\rho
_{c}F+p_{c}F\right) .  \label{FMfai2}
\end{equation}%
where the density and pressure of scalar field $\phi $ are%
\begin{align*}
\rho _{f}& =\frac{1}{2}\dot{\phi}^{2}+V\left( \phi \right) \\
p_{f}& =\frac{1}{2}\dot{\phi}^{2}-V\left( \phi \right) ,
\end{align*}%
and $\rho _{e}\equiv \rho _{c}F$ ($p_{e}\equiv p_{c}F$) may be understood as
effective density (pressure) of the energy component in scalar-tensor
theory:
\begin{equation*}
\rho _{c}=-\frac{n}{8\pi GF}H\dot{F}
\end{equation*}%
\begin{equation*}
p_{c}=\frac{1}{8\pi GF}\left[ \ddot{F}+\left( n-1\right) H\dot{F}\right] .
\end{equation*}%
Obviously, we need to consider the non-equilibrium thermodynamics. We select
$\rho _{i}=(\rho _{f},F,\rho _{e})$. Now recalling the continuous equation (%
\ref{Con1}) and using the first Friedmann equation (\ref{FMfr1}), one can
find that the l.h.s in Eq. (\ref{con}) reads
\begin{align}
\dot{\rho}_{eff}& =\frac{n(n-1)}{16\pi G}\left( \frac{\partial H^{2}}{%
\partial \rho }\dot{\rho}+\frac{\partial H^{2}}{\partial \rho _{i}}\dot{\rho}%
_{i}\right)  \notag \\
& =\frac{n(n-1)}{16\pi G}\left[ \frac{16\pi G}{n(n-1)}\frac{1}{F}\dot{\rho}+%
\frac{\partial H^{2}}{\partial \rho _{i}}\dot{\rho}_{i}\right]  \notag \\
& =\frac{1}{F}\dot{\rho}+\frac{n(n-1)}{16\pi G}\frac{\partial H^{2}}{%
\partial \rho _{i}}\dot{\rho}_{i}  \label{r3}
\end{align}%
while the r.h.s in Eq. (\ref{con}) reads by the second Friedmann equation (%
\ref{FMfai2})%
\begin{equation}
-nH(\rho _{eff}+p_{eff})=-nH\frac{1}{F}\left( \rho +p+\rho _{f}+p_{f}+\rho
_{e}+p_{e}\right) .  \label{r4}
\end{equation}%
Employing the continuous equation (\ref{Con2}) to Eqs. (\ref{r3}) and (\ref%
{r4}), one can find%
\begin{equation}
\frac{\partial H^{2}}{\partial \rho _{i}}\dot{\rho}_{i}=-\frac{16\pi G}{(n-1)%
}\frac{1}{F}\left[ H\left( \rho _{f}+p_{f}+\rho _{e}+p_{e}\right) +q\right] .
\label{ls1}
\end{equation}%
By substituting Eq. (\ref{ls1}) into Eq. (\ref{dE}), we obtain%
\begin{align}
dE& =nVH(\rho +p)dt-Vqdt-Vdt\frac{1}{\frac{\partial H^{2}}{\partial \rho }}%
\frac{\partial H^{2}}{\partial \rho _{i}}\dot{\rho}_{i}  \notag \\
& =nVH(\rho +p+\rho _{f}+p_{f})dt+nVH\left( \rho _{e}+p_{e}\right) dt,
\label{frdE1}
\end{align}%
where $nVH\left( \rho _{e}+p_{e}\right) dt$ denotes the energy flux of
effective energy component $\rho _{e}$ in scalar-tensor gravity. Then we
have the first law%
\begin{equation*}
TdS+Td_{i}S=dE,
\end{equation*}%
where the entropy production and entropy can be obtained from Eq. (\ref{dis}%
) and Eq. (\ref{S}) respectively
\begin{equation*}
d_{i}S=-S_{E}dF,
\end{equation*}%
and%
\begin{equation}
S=S_{E}F=\frac{A}{4G}F.  \label{S st}
\end{equation}

\bigskip Setting $f_{2}=f_{3}=0$, the mass-like function (\ref{mbar1}) for $%
n=3$ is%
\begin{equation}
\tilde{M}=\frac{n(n-1)\Omega_{n}\tilde{r}_{A}^{n-2}}{16\pi G}(1+h^{ab}\tilde{%
r}_{,a}\tilde{r}_{,b})F.  \label{mbar st}
\end{equation}

It is interesting to note that entropy expressions (\ref{S lovelock}), (\ref%
{Sfr}), and (\ref{S st}) are consistent with black hole entropy of Lovelock
gravity \cite{Cai1}, nonlinear gravity \cite{Wald} and scalar-tensor gravity
\cite{Cai2}, respectively. The obtained mass-like functions (\ref{mbar love}%
), (\ref{mbar fr}), and (\ref{mbar st}) also agree to those presented in
\cite{Gong}.

\section{Thermodynamics of Randall-Sundrum braneworld with nontrivial bulk}

\bigskip We consider a $n$-dimensional brane embedded in a ($n+2$%
)-dimensional space-time. For convenience and without loss of generality we
choose the extra dimension along the coordinates $y$ such that the brane is
located at $y=0$ and the bulk has $Z_{2}$ symmetry under the transformation $%
y\rightarrow-y$. The action is%
\begin{equation}
S=\frac{1}{2\kappa_{5}^{2}}\int d^{n+2}x\sqrt{-g}\{\mathcal{L}_{EH}\}+\int
d^{n+2}x\sqrt{-g}\{\mathcal{L}_{M}\}+\int d^{n+1}x\sqrt{-\tilde{g}}\{%
\mathcal{L}_{m}-2\lambda\},  \label{Sgrav}
\end{equation}
where $g$ ($\tilde{g}$), $\kappa_{n+2}$ ($\kappa_{n+1}$), and $\mathcal{L}%
_{M}$ ($\mathcal{L}_{m}$) are the bulk (brane) metric, bulk (brane)
gravitational constant, and bulk (brane) matter fields, respectively. $%
\mathcal{L}_{EH}=R-2\Lambda$ is the five-dimensional Einstein-Hilbert
Lagrangian with negative cosmological constant $\Lambda<0$. $\lambda$ is the
brane tension (or cosmological constant). For convenience, we will choose
the unit that $\kappa_{n+2}=1$.

\bigskip By varying the action in Eq. (\ref{Sgrav}) with respect to the bulk
metric, we obtain the field equation%
\begin{equation}
G_{AB}=\left. T_{AB}\right\vert _{total},  \label{field equation}
\end{equation}
The total energy-momentum tensor $\left. T_{AB}\right\vert _{total}$ is
decomposed into bulk and brane components%
\begin{equation*}
\left. T_{AB}\right\vert _{total}=\left. T_{AB}\right\vert _{bulk}+\left.
T_{AB}\right\vert _{brane}\hat{\delta}(y).
\end{equation*}
Here, we use the normalized Dirac delta function, $\hat{\delta}(y)=\sqrt {%
\tilde{g}/g}\delta(y)$. The bulk component is
\begin{equation*}
\left. T_{AB}\right\vert _{bulk}=-\Lambda g_{AB}+T_{AB},
\end{equation*}
where $T_{AB}$ denotes all possible energy-momentum in the bulk. The brane
component is written as
\begin{equation*}
\left. T_{AB}\right\vert _{brane}=-\lambda\tilde{g}_{AB}+\tilde{T}_{AB},
\end{equation*}
where energy momentum tensor $\tilde{T}_{AB}$ represents matter on the brane
with energy density $\rho$ and pressure $p$.

The ($n+2$)-dimensional line element in the bulk is given by
\begin{equation*}
ds^{2}=-n^{2}(t,y)dt^{2}+a^{2}(t,y)\gamma_{ij}dx^{i}dx^{j}+b^{2}(t,y)dy^{2}
\end{equation*}
where $\gamma_{ij}$ is a $n$-dimensional maximally symmetric metric whose
spatial curvature is characterized by $k=0,\pm1$. Here we are interested in
spatially flat brane $k=0$. We choose the coefficients $n(t,0)=1$ so that $t$
is the proper time along the brane. For simplicity, we assume that the fifth
dimension is static $\dot{b}=0$ and we set $b=1$.

To determine the Friedmann equations on the brane, we need the junction
condition on the brane. For convenience, we use prime and dot to denote the
derivative with respect to $y$ and $t$, respectively. The jump of the $(00)$
and $(ij)$ components of the field equation (\ref{field equation}) across
the brane gives%
\begin{equation}
\frac{a_{+}^{\prime}}{a_{0}}=-\frac{1}{2n}(\rho+\lambda),  \label{ap}
\end{equation}%
\begin{equation}
\frac{n_{+}^{\prime}}{n_{0}}=\frac{p+\frac{n-1}{n}\rho}{2},  \label{np}
\end{equation}
where $2a_{+}^{\prime}=-2a_{-}^{\prime}$ and $2n_{+}^{\prime}=-2n_{-}^{%
\prime }$ are the discontinuities of the first derivatives. Substituting Eq.
(\ref{ap}) and Eq. (\ref{np}) into the $(05)$ component of the field
equation (\ref{field equation})%
\begin{equation}
n(\frac{n^{\prime}}{n}\frac{\dot{a}}{a}-\frac{\dot{a}^{\prime}}{a})=T_{05},
\label{05}
\end{equation}
we obtain the continuity equation
\begin{equation}
\dot{\rho}+nH(\rho+p)=2T_{05}.  \label{ro}
\end{equation}
From the $(00$) and $(55)$ equations, following \cite{Binetruy}, we find a
set of functions $a(t,y)$ and $n(t,y)$
\begin{equation}
\Phi=\left( \frac{\dot{a}}{na}\right) ^{2}-\frac{a^{\prime2}}{a^{2}}-\frac{%
2\Lambda_{n+2}}{n(n+1)}.  \label{Fai}
\end{equation}
When the bulk is Ads$_{5}$, it satisfies the constraint equation%
\begin{equation*}
\Phi=0.
\end{equation*}
But for nontrivial bulk, $\Phi$ will not be a constant. Even when $T_{05}=$ $%
T_{55}=0$, a weyl radiation term$\sim a^{-n-1}$ may appear. One can find the
evolved equation of $\Phi$ in \cite{Wu}, and the physical meaning of $\Phi$
is the correction to the bulk cosmological constant. Substituting the
junction conditions (\ref{ap}) and (\ref{np}) into Eq. (\ref{Fai}), we can
obtain the first Friedmann equation%
\begin{equation}
H^{2}=\frac{1}{4n^{2}}\rho^{2}+\frac{1}{2n^{2}}\lambda\rho+\Phi
\label{FM1RS}
\end{equation}
where the Randall-Sundrum fine-turning condition%
\begin{equation*}
\frac{1}{4n^{2}}\lambda^{2}+\frac{2\Lambda_{n+2}}{n(n+1)}=0
\end{equation*}
has been used.

There are other freedoms besides $\rho $ in the first Friedmann equation (%
\ref{FM1RS}), so one should use the entropy expression (\ref{S}). We select $%
\rho _{1}=\Phi $. Now Eq. (\ref{dE}) reads%
\begin{align*}
dE& =nVH(\rho +p)dt-Vqdt-Vdt\frac{1}{\frac{\partial H^{2}}{\partial \rho }}%
\frac{\partial H^{2}}{\partial \rho _{i}}\dot{\rho}_{i} \\
& =\left[ nVH(\rho +p)dt-2VT_{05}dt\right] -Vdt\frac{2n^{2}\tilde{r}_{A}}{%
\sqrt{\tilde{r}_{A}^{2}\lambda ^{2}-4n^{2}\left( \tilde{r}_{A}^{2}\Phi
-1\right) }}\dot{\Phi},
\end{align*}%
Then we have the first law%
\begin{equation*}
TdS+Td_{i}S=dE,
\end{equation*}%
where the entropy production and entropy can be obtained from Eq. (\ref{dis}%
) and Eq. (\ref{S}) respectively
\begin{equation*}
d_{i}S=\frac{8\pi n^{2}\tilde{r}_{A}^{n-1}\Omega _{n}}{(n-1)}d\frac{\tilde{r}%
_{A}}{\sqrt{\tilde{r}_{A}^{2}\lambda ^{2}-4n^{2}\left( \tilde{r}_{A}^{2}\Phi
-1\right) }}
\end{equation*}%
and%
\begin{equation}
S=\frac{8\pi n^{2}\Omega _{n}\tilde{r}_{A}^{n}}{(n-1)\sqrt{\lambda
^{2}-4n^{2}\tilde{r}_{A}^{2}\left( \Phi -1\right) }}.  \label{SFai}
\end{equation}%
The mass-like function (\ref{mbar1}) for $n=3$ is now expressed as%
\begin{equation*}
\tilde{M}=\frac{16\pi G}{n(n-1)}\frac{1}{\frac{\partial H^{2}}{\partial \rho
}}M=\frac{2n^{2}\Omega _{n}\tilde{r}_{A}^{n-1}}{\sqrt{\tilde{r}%
_{A}^{2}\lambda ^{2}-4n^{2}\left( \tilde{r}_{A}^{2}\Phi -1\right) }}(1+h^{ab}%
\tilde{r}_{,a}\tilde{r}_{,b}).
\end{equation*}%
To compare with the result obtained in \cite{Cao1}, we set $\Phi =0$. Now
the entropy expression (\ref{S0}) should be used,%
\begin{equation}
S=\int \frac{4\pi \tilde{r}_{A}^{n-2}\Omega _{n}}{\frac{\partial H^{2}}{%
\partial \rho }(\tilde{r}_{A})}d\tilde{r}_{A}=\int \frac{8\pi n^{2}\Omega
_{n}\tilde{r}_{A}^{n-1}}{\sqrt{\tilde{r}_{A}^{2}\lambda ^{2}+4n^{2}}}d\tilde{%
r}_{A}.  \label{57}
\end{equation}%
One can find that the entropy expression is consistant with the result
obtained in \cite{Cao1}. The corresponding first law (\ref{first law1}) reads%
\begin{equation*}
TdS=dE=-V\dot{\rho}dt=n\Omega _{n}\tilde{r}_{A}^{n}H(\rho +p),
\end{equation*}%
where Eq. (\ref{dE}) without $q$ and $\rho _{i}$ has been used to get the
second equality.

Setting $f_{2}=f_{3}=0$, the mass-like function (\ref{mbar}) now reads%
\begin{equation*}
\tilde{M}=\frac{16\pi G}{n(n-1)}\frac{1}{\frac{\partial H^{2}}{\partial\rho}}%
M=\frac{2n^{2}\Omega_{n}^{2}\tilde{r}_{A}^{2n-3}}{\sqrt{\tilde{r}%
_{A}^{2}\lambda^{2}+4n^{2}}}(1+h^{ab}\tilde{r}_{,a}\tilde{r}_{,b}).
\end{equation*}

\section{Conclusions and discussions}

In this paper, we have constructed the first law of thermodynamics on the
apparent horizon of generalized gravity theories, including Einstein
gravity, Lovelock, $f(R)$ and scalar-tensor gravity theories. We have also
generalized our study to the Randall-Sundrum braneworld with nontrivial bulk
and obtained the corresponding entropy and the first law of thermodynamics.
It is interesting to observe that entropy expressions (\ref{S0}) and (\ref{S}%
) are general, which can lead to the consistency with the black hole entropy
obtained in extended gravity theories such as Lovelock, $f(R)$ and
scalar-tensor gravity theories. This gives us the hope that the general
entropy expression can be used to shed some lights on even more generalized
gravity theory such as the Randall-Sundrum braneworld etc., where until now
there is no exact black hole solution obtained.

One may argue that there seems to be ambiguities in the entropy expressions
(12) (18), since one might add proper quantity to the expression of entropy,
which may vanish in the Einstein gravity, and this extra term could be
absorbed into the redefinition of the entropy production. This worry is not
necessary, since the known black hole entropy in different gravity theories
will strictly restrict the form of the additional quantities in the entropy
expressions. Entropy expressions (27) (38) and (46) are consistent with the
known black hole entropy in Lovelock gravity [21], nonlinear gravity [22]
and scalar-tensor gravity [23], respectively. This fact shows that it is not
needed to add additional quantities to the entropy expressions. For the
braneworld case, without adding additional terms, (57) reduces to the area
formula in $n+1$ dimensions in the large horizon limit, while in the small
horizon limit, it becomes the area formula in the bulk [10,27]. On the other
hand, we have built the relation between the general mass-like functions and
the entropy expressions. The obtained mass-like functions (28) (39) and (47)
are in agreement with those presented in [18]. The general mass-like
functions have dimension of energy. However if one adds other quantities in
(12) (18), the derived mass-like function cannot reduce to that obtained in
[18]. This serves as another restriction on adding additional terms to the
entropy expressions.

Our formalism of constructing the first law of thermodynamics is general and
can be applied to any gravity theory no matter matter contents are conserved
or not. We find that the non-equilibrium entropy production appears due to
the other dynamic fields besides the ordinary matter dominating the
cosmological evolution.

We have argued that the mass-like function presented in \cite{Gong} is
general in extended gravity theories. In Lovelock gravity, the conjectured
generalized Misner-Sharp mass in \cite{Maeda1} is the special case of the
mass-like function. This sparks us to further investigate the physical
meaning of the mass-like function and its relation to the generalized
Misner-Sharp mass in generalized gravity theories.

\begin{acknowledgments}
This work was partially supported by the NSFC, Shanghai Education
Commission, Science and Technology Commission. S. F. Wu and G. H. Yang were
also supported by the NSFC under Grant No. 10575068, the Shanghai Education
Development Foundation, and the Natural Science Foundation of Shanghai
Municipal Science Technology Commission under grant Nos. 04ZR14059 and
04dz05905.
\end{acknowledgments}


\begin{thebibliography}{99}
\bibitem{Jacobson} T. Jacobson, Phys. Rev. Lett. \textbf{75} (1995) 1260.

\bibitem{Padmanabhan} T. Padmanabhan, Class. Quant. Grav. \textbf{19} (2002)
5387; Phys. Rep. \textbf{406} (2005) 49.

\bibitem{Frolov} A. V. Frolov, L. Kofman, JCAP \textbf{0305} (2003) 009.

\bibitem{Danielsson} U. H. Danielsson, Phys. Rev. \textbf{D 71} (2005)
023516.

\bibitem{Bousso} R. Bousso, Phys. Rev. \textbf{D 71} (2005) 064024.

\bibitem{Cai} R. G. Cai, S. P. Kim, JHEP \textbf{0502} (2005) 050.

\bibitem{Eling} C. Eling, R. Guedens, T. Jacobson, Phys. Rev. Lett. \textbf{%
96} (2006) 121301.

\bibitem{Akbar} M. Akbar, and R. G. Cai, Phys. Lett. \textbf{B 64}8 (2007)
243.

\bibitem{Cao} R. G. Cai, L. M. Cao, Phys. Rev. \textbf{D 75} (2007) 064008.

\bibitem{Cao1} R. G. Cai, L. M. Cao, Nucl. Phys. \textbf{B 785} (2007) 135.

\bibitem{Sheykhi} A. Sheykhi, B. Wang, and R. G. Cai, Nucl. Phys. \textbf{B
779} (2007) 1.

\bibitem{Sheykhi1} A. Sheykhi, B. Wang, and R. G. Cai, Phys. Rev. \textbf{D
76} (2007) 023515.

\bibitem{Wu1} S. F. Wu, G. H. Yang, and P. M. Zhang, arXiv:0710.5394.

\bibitem{Cai3} R. G. Cai, Y. Gong and B. Wang, JCAP \textbf{0603} (2006) 006.

\bibitem{Apostolopoulos} P. S. Apostolopoulos and N. Tetradis, Phys. Rev.
\textbf{D 74} (2006) 064021.

\bibitem{wang1} S. Y. Yin, B. Wang, E. Abdalla, C. Y. Lin, arXiv:0708.0992.

\bibitem{wang2} A. Sheykhi, B. Wang, N. Riazi, Phys. Rev. \textbf{D 75}
(2007) 123513.

\bibitem{Gong} Y. Gong, A. Wang, Phys. Rev. Lett. \textbf{99} (2007) 211301.

\bibitem{Maeda} H. Maeda, Phys. Rev. \textbf{D 73} (2006) 104004.

\bibitem{Maeda1} H. Maeda and M. Nozawa, arXiv:0709.1199.

\bibitem{Cai1} R. G. Cai, Phys. Lett. \textbf{B 582} (2004) 237.

\bibitem{Wald} R. M. Wald, Phys. Rev. \textbf{D 48} (1993) 3427; G. Cognola,
E. Elizalde, S. Nojiri, S. D. Odintsov, S. Zerrbini, JCAP \textbf{0502}
(2005) 010; I. Brevik, S. Nojiri, S.D. Odintsov, L. Vanzo, Phys. Rev.
\textbf{D 70} (2004) 043520.

\bibitem{Cai2} R. G. Cai and Y. S. Myung, Phys. Rev. \textbf{D 56} (1997)
3466.

\bibitem{Lovelock} D. Lovelock, J. Math. Phys. \textbf{12} (1971) 498.

\bibitem{Binetruy} P. Binetruy, C. Deffayet and D. Langlois, Nucl. Phys.
\textbf{B 565} (2000) 269; P. Binetruy, C. Deffayet, U. Ellwanger and D.
Langlois, Phys. Lett. \textbf{B 477} (2000) 285.

\bibitem{Wu} S. F. Wu, A. Chatrabhuti, G. H Yang, and P. M. Zhang, Phys.
Lett. \textbf{B} \textbf{659} (2007) 45.
\end{thebibliography}
\end{document}